\documentclass[10pt]{article}
\usepackage{amsmath}
\usepackage{graphicx}
\usepackage{amssymb}
\usepackage{amsfonts}
\usepackage{latexsym}
\usepackage[pdftex,pdfmenubar=true,bookmarks=true,pdftoolbar=true]{hyperref}
\hypersetup{colorlinks,citecolor=green,filecolor=magenta,linkcolor=red,urlcolor=cyan,pdftex}
\begin{document}

\begin{center}
\begin{flushright}\begin{small}    UFES 2012
\end{small} \end{flushright} \vspace{1.5cm}
\huge{Thermodynamics of phantom black holes in Einstein-Maxwell-Dilaton theory}
\end{center}

\begin{center}
{\small \bf Manuel E. Rodrigues $^{(a)}$}\footnote{E-mail
address: esialg@gmail.com}\ and
{\small \bf Zui A. A. Oporto$^{(a)}$}\footnote{E-mail address:
azurnasirpal@gmail.com}
 \vskip 4mm

(a)  \ Universidade Federal do Esp\'{\i}rito Santo \\
Centro de Ci\^{e}ncias Exatas - Departamento de F\'{\i}sica\\
Av. Fernando Ferrari s/n - Campus de Goiabeiras\\ CEP29075-910 -
Vit\'{o}ria/ES, Brazil \vskip 2mm

\end{center}

\begin{abstract}

A thermodynamic analysis of the black hole solutions coming from the Einstein-Maxwell-Dilaton theory (EMD) in $4$D is done. By consider the canonical and grand-canonical ensemble, we apply standard method as well as a recent method known as Geometrothermodynamics (GTD). We are particularly interested in the characteristics of the so called phantom black hole solutions. We will analyze the thermodynamics of these solutions, the points of phase transition and their extremal limit. Also the thermodynamic stability is analyzed. We obtain a mismatch of the between the results of the GTD method when compared with  the ones obtained by the specific heat, revealing a weakness of the method, as well as possible limitations of its applicability to very pathological thermodynamic systems. We also found that normal and phantom solutions are locally and globally unstable, unless for certain values of the coupled constant of the EMD action. We also shown that the anti-Reissner-Nordstrom solution does not posses extremal limit nor phase transition points, contrary to the Reissner-Nordstrom case.
\end{abstract}

Pacs numbers: 04.70.-s; 04.20.Jb; 04.70.Dy. 


\section{Introduction}
\hspace{0,6cm}
Since the discovery made by Hawking \cite{hawking1} of the thermodynamic properties  of the black holes related to the quantum phenomena, many were the interests on studying the properties of various kinds of solutions obtained from General Relativity and their modifications. In close analogy to the usual thermodynamics, the black hole thermodynamics is based upon four basic principles, the zero law and the other three laws, which are all analogous to the usual ones \cite{bardeen,lousto}. Therefore, a new black hole solution can be physically interpreted also by the analysis of their thermodynamic properties. Furthermore, we can analyse the thermodynamic stability of a new solution through its properties.
\par
There are several methods to study the thermodynamics properties and stability of a black hole. A basic reference on this is the work of Davies \cite{davies}. We can mention other more recent methods, such as the Geometrothermodynamics \cite{quevedo}, and the Hamiltonian thermodynamics \cite{louko}. For the present work the first two methods will developed, always in parallel so we can compare their results. We want to stress that the main goal of this paper is to analyze the thermodynamics properties and stability of the solutions known as phantom black holes, specifically those coming from the Einstein-(anti)Maxwell-(anti)Dilaton (EMD) theory. These solutions come from the minimal coupling of the Einstein-Hilbert action with a scalar field, that could be either dilatonic or phantom, which at the same time this is coupled with Maxwell field, that can be a spin-$1$ normal or phantom field. The phantom term furnishes the contribution of negative energy density, which justify the nomenclature.
\par
In order to develop the analysis of the thermodynamic properties of this class of black holes, let us briefly illustrate the interest to studying phantom solutions in black hole physics. The programs of evolution of our universe, specially the ones for the spectrum of anisotropies of the cosmic background radiation on the one hand and for the relation magnitude versus red-shift of the supernovae type Ia on the other, have pointed out today an accelerated expansion of the universe, dominated by an exotic fluid of negative pressure. Furthermore, there are evidences suggesting this exotic fluid could be of phantom nature \cite{hannestad}. Hence, several classes of black holes (also wormholes, see \cite{kirill1}) have been found which have phantom characteristics. An important class of phantom solutions is the Gibbons and Rasheed's of the EMD theory \cite{gibbons}. Several others generalizations were obtained, such as the higher-dimensional black holes by Gao and Zhang \cite{gao} and the higher-dimensional black branes by Grojean et al \cite{grojean}. The analysis of the algebra produced by a metric with two times in higher dimensions, which provides phantom fields in 4D, was developed by Hull \cite{hull} and for Sigma models by Clement et al \cite{gerard2}. On this work, we will study some solutions coming from the EMD theory, which were studied in detail in \cite{gerard1}.
\par
There are some methods of analysis in black hole thermodynamics theory dubbed as geometrical, because they make use of differential geometry to determine thermodynamic properties such as: points at which black holes become extremal or they pass through a phase transition and thermodynamic stability  of the system. One of the first methods was proposed by Rao \cite{rao}, subsequently developed by other authors \cite{amari}. Later, the work of Weinhold \cite{weinhold} and Ruppeiner \cite{ruppeiner} were frequently used for the study of the black hole thermodynamics. The method we will explore in this work is known as Geometrothermodynamics (GTD), however,  the results obtained with this method will always be compared with the ones obtained by the usual non-geometric methods. The GTD had been shown to be equivalent, and in some situations, even superior in many aspects when compared with the usual non-geometric approaches. The GTD has been widely used in the literature to study the most diverse classes  of black holes \cite{quevedo1,quevedo5}. The results obtained by the GTD methods reconciles some inconsistencies between the Weihold and Ruppeiner methods, for example, we can mention some cases in which a black hole suffer a phase transition according to one method but not by other one. Yet, there is still a good concordance when different thermodynamic potentials are chosen, like for example the mass and entropy representations, this in virtue of the invariance of the formalism by Legendre transformations. Finally, we have mention that the results are independent of the particular thermodynamic ensemble considered.
\par
However, we also have to point out that the GTD method can contain some inconsistencies when compared with the more usual analysis done by the specific heat. Recently it has shown that for the cases Reissner-Nordstrom-AdS and (phantom case) anti-Reissner-Nordstrom-AdS black holes \cite{deborah}, the GTD method does not reproduce the results obtained by the specific heat method. We will arrive to the same conclusion here in the case of phantom black holes within the EMD theory. 
\par
This paper is organized as follows: In Section \ref{sec2} a summary of the static and spherically symmetric solution for the EMD theory in $4$D\footnote{From here it will be implied that the black hole solutions discussed here are static and spherically symmetric solution coming from the EMD theory in $4$D.} and the derivation of thermodynamics variables are presented.  In Section \ref{sec3} the GTD  method is reviewed in some detail. Section \ref{sec4} is divided in two parts, the GTD method is applied to the classes of asymptotically flat black holes of the EMD theory and in particular to anti-Reissner-Nordstrom case in sub-section \ref{sbsec4.1}, and  the analysis of the local and global stability is developed in sub-section \ref{sbsec4.2}. Conclusions and perspectives of the work are presented in Section \ref{sec5}.



\section{ Phantom black hole solutions and the first law of thermodynamics}\label{sec2}

\hspace{0,6 cm} In this section we present the class of solutions coming from Einstein-Maxwell-Dilaton theory, their relevant parameters, the formulation of the first law of thermodynamics, as well as the fundamental ingredients necessaries for a  detailed analysis of the thermodynamics properties of these solutions.
\par
We begin by defining the Einstein-Maxwell-Dilaton (EMD) action as
\begin{equation}
S=\int dx^{4}\sqrt{-g}\left[  \mathcal{R}-2\,\eta_{1}
g^{\mu\nu}\nabla_{\mu}\varphi\nabla_{\nu }\varphi+\eta_{2} \,e^{
2\lambda\varphi}F^{\mu\nu}F_{\mu\nu}\right]  \label{action1}\; ,
\end{equation}
where the first term corresponds to the usual Einstein-Hilbert action, the second one is the kinetic term of the scalar field (dilatonic or phantom) and the third one is the coupling term between the scalar and Maxwell fields, with real valued coupling constant $\lambda$. The coupling constant $\eta_1$ takes the values
$\eta_1=+1$ (dilaton) or $\eta_1=-1$ (anti-dilaton). The parameter $\eta_2$ can be $\eta_2=+1$ (Maxwell) or $\eta_2=-1$ (anti-Maxwell). Depending on whether the contribution of the  energy density is positive or negative, which is determined by $\eta_{1,2}$, the kinetic term of the scalar field and the coupling term with the Maxwell field can be normal (ordinary) or phantom.

Now, we use some well established results about the class of black hole solutions in EMD theory. According to \cite{gerard1}, Eq. $(2.19)$ in Section II, let us choose the solutions of $\omega(u)$ and $J(u)$ as the function $\sinh (u)$, where the horizon is non-degenerate and $u_0>0$, and let us perform a reparametrization of the radial coordinate as 
\begin{equation}\label{1.1}
u=\frac{1}{(r_{+}-r_{-})}\ln \left( \frac{f_{+}}{f_{-}}\right)\; ,\qquad
f_{\pm}=1- \frac{r_{\pm}}{r}\; ,
\end{equation}
with
\begin{equation}\label{1}
r_\pm = \pm \frac{2a}{1 - e^{\mp 2au_{0}}} \qquad (r_{+}-r_{-} = 2a)\;,
\end{equation}
then we obtain the solution
\begin{eqnarray}
dS^{2}&=&f_{+}f_{-}^{\gamma}dt^{2}
-f_{+}^{-1}f_{-}^{-\gamma}dr^{2}
-r^{2}f_{-}^{1-\gamma}d\Omega^{2} \;  ,\label{2} \\
F&=&-\frac{q}{r^2}dr\wedge dt\; ,\;
e^{-2\lambda\varphi}=f_{-}^{1-\gamma}\label{3}
\; ,
\end{eqnarray}
where $\varphi_0=0$, $\gamma=\lambda_{-}/\lambda_{+}$ (for $\eta_{1}=1$, $-1<\gamma<1$, and $\eta_{1}=-1, \gamma\in (-\infty,-1)$ $\cup$ $(1,+\infty)$), $0<r_{-}<r_{+}$ for $\eta_2\lambda_{+}>0$, and finally $r_{-}<0<r_{+}$ for $\eta_2\lambda_{+}<0$. This is the exact solution of a spherically symmetric black hole, asymptotically flat, electrically charged and static, with internal horizon ``$r_{-}$"\footnote{A detailed discussion of the causal structure for the phantom case can be found in \cite{gerard1}.} and event horizon ``$r_{+}$", which is related to the physical parameters, mass and charge of the black hole, trough the relations
\begin{align}
M &  =\frac{r_{+}+\gamma r_{-}}{2}\label{4}\; ,\\
q &  =\sqrt{\frac{1+\gamma}{2}}\sqrt{\eta_{2}r_{+}r_{-}}\;
.\label{5}
\end{align}
\par
Now, we are interested in the geometrical analysis representing semi-classical gravitational effects of the black hole solutions mentioned before. By semi-classical we mean quantize the called matter fields, while leaving classical the background gravitational field. Therefore we will work with the semi-classic thermodynamics of black holes, studied first by Hawking \cite{hawking1}, and further developed by many other authors \cite{davies2}.

There are several techniques to derive the Hawking temperature law. For example we can mention the Bogoliubov coefficients \cite{ford} and the energy-momentum tensor methods \cite{davies,davies2}, by the euclidianization of the metric \cite{hawking2}, the transmission and reflection coefficients \cite{gerard3,kanti}, the analysis of the anomaly term \cite{robinson}, and by the black hole superficial gravity \cite{jacobson}. Since all these methods have been proved to be equivalents \cite{manuel}, then we opt, without loss of generality, to calculate the Hawking temperature by the superficial gravity method.

The superficial gravity of a black hole is given by the expression \cite{wald}:
\begin{equation}
\kappa =\left[  \frac{g_{00}^{\prime}}{2\sqrt{-g_{00}g_{11}}}\right]_{r\;=\;r_{+}}\, ,\label{10}
\end{equation}
where $r_{+}\;$ is the radius of the event horizon. The relationship between the Hawking temperature of the black hole and its superficial gravity is given by the formula \cite{hawking1,jacobson}:
\begin{equation}
T=\frac{\kappa}{2\pi}\, .\label{11}
\end{equation}

Therefore, for the the black hole defined by (\ref{2}), we have that the superficial gravity (\ref{10}) takes the form
\begin{equation}
 \kappa =\frac{\left(  r_{+}-r_{-}\right)  ^{\gamma}}{2r_{+}^{1+\gamma}}\, ,\label{12}
\end{equation}
and the corresponding Hawking temperature (\ref{11}) would be
\begin{equation}
T=\frac{\left(  r_{+}-r_{-}\right)  ^{\gamma}}{4\pi r_{+}^{1+\gamma}}\, .\label{th1}
\end{equation}

We define the area of the horizon of the black hole as
\begin{equation}
A=\int _{0}^{2\pi}\int_{0}^{\pi}\sqrt{g_{22}g_{33}}\;d\theta d\phi\Big|_{r=r_{+}}=4\pi r^2f_{-}^{1-\gamma}\Big|_{r=r_+}=4\pi r_{+}^{1+\gamma}\left(r_{+}-r_{-}\right)^{1-\gamma}\label{a1}\;.
\end{equation}
Then the entropy of the black hole can be defined as \cite{bardeen}
\begin{equation}
S=\frac{1}{4}A=\pi r_{+}^{1+\gamma}\left(r_{+}-r_{-}\right)^{1-\gamma}\label{s1}\;.
\end{equation}
On the other hand, working out (\ref{3}) we obtain the electric potential at the event horizon which reads
\begin{equation}
A_{0}=\int_{+\infty}^{r}F_{10}(r^{\prime})dr^{\prime}\Big|_{r=r_+}=\frac{q}{r_{+}}\label{pot1}\;.
\end{equation}

Using eqs. (\ref{4}), (\ref{5}) and (\ref{s1}), we can write the differential forms of the mass, charge and entropy:
\begin{eqnarray}
\label{dif} \left\{\begin{array}{lr}
dM=\frac{1}{2}\left(dr_{+}+\gamma dr_{-}\right)\; ,\\
dq=\eta_{2}\left(\frac{1+\gamma}{4q}\right)\left(r_{-}dr_{+}+r_{+}dr_{-}\right)\;,\\
dS=\frac{\pi r_{+}^{1+\gamma}}{(r_{+}-r_{-})^{\gamma}}\left\{ \left[2-(1+\gamma)\frac{r_{-}}{r_{+}}\right]dr_{+}-(1-\gamma)dr_{-}\right\}\;,
\end{array}\right.
\end{eqnarray}
these relations, together with (\ref{th1}) and (\ref{pot1}), satisfy the first law of black hole thermodynamics \cite{bardeen}
\begin{equation}
dM=TdS+\eta_{2}A_{0}dq\label{plt}\;.
\end{equation}
Note that when $\eta_{2}=-1$, the first law is generalized for the Einstein-anti-Maxwell-Dilaton sector, where the second term, which is related to the work, suffers a change of signal as a consequence of a negative energy contribution to the system. The exact formula of Eq. (\ref{plt}), originally known as the Smarr formula \cite{smarr}, can be integrated resulting in
\begin{equation}
M=2TS+\eta_{2}A_{0}q\label{smarr}\;.
\end{equation}
In this case we also have a change of signal of the work term for the phantom case.

We can write the horizons in terms of the mass and charge parameters of the black hole as follows:
\begin{eqnarray}
r_{+}&=&M+\sqrt{M^2-\frac{2\eta_{2}\gamma q^2}{(1+\gamma)}}\label{r1}\;,\\
r_{-}&=&\frac{1}{\gamma}\left(M-\sqrt{M^2-\frac{2\eta_{2}\gamma q^2}{(1+\gamma)}}\right)\label{r2}\;.
\end{eqnarray}

It is important to note that the possible extremal case, i.e., $r_{+}=r_{-}$, have to be analyzed carefully, because according to the causal structure described in \cite{gerard1}, for the ordinary EMD theory $r_{-}$ is a curvature singularity, and for the phantom case in which $\eta_2\lambda_+<0$ ($r_{-}<0$) is a true singularity . Hence, in the last case we have a causal structure identical to the Schwarzschild solution with just one horizon. In some phantom cases where $\gamma$ takes discrete (integer) values, we have two horizons that can be crossed where the radial coordinate $r$ for the internal horizon is no longer $r_{-}$, but $r=0$, with $r_{-}$ being a singularity where the geodesics ends. When this is so, it is not simple to determine if the extremal limit exist, because $r_{+}\rightarrow 0$ represents a regime which has not yet been analyzed in the literature. The analysis by the Geometrothermodynamics method could provide a new insight to understand this pathological solution, but we will see that, just as the usual method using the specific heat, some subtleties are still unavoidable. Maybe this subtleties can be understood only within the context of more fundamental quantum analysis.
\par
When the normal EMD theory is considered, the extremal limit still provides a structure quite similar to what is called a naked singularity; classified as a light-like naked singularity, which can be reached only after an infinity lapse of time. A semi-classical analysis of such a structure, i.e., a non-asymptotically flat black hole, has been studied in \cite{gerard3}, but the analysis  does not contain a conclusive result in relation to the thermodynamic properties. So once again it seems like some fundamental quantum theory of gravity should be considered at this point to depth into this question.
\par
We have briefly discussed here the causal structure and the extremal limit of this class of solutions, now we are interested in their thermodynamic properties, which we will see present subtleties precisely for the particular cases considered in this work.
\par
Taking the eqs. (\ref{s1}), (\ref{pot1}), (\ref{r1}) and (\ref{r2}), we can rewrite the temperature, entropy and the electric potential as functions of the mass and electric charge as follows:
\begin{eqnarray}
T(M,q)&=&\frac{1}{4\pi\gamma^{\gamma}}\left(M+\sqrt{M^2-\frac{2
\eta_{2}\gamma q^2}{ (1+\gamma)}}\right)^{1+\gamma}\times\nonumber\\
\end{eqnarray}
\begin{eqnarray}
&&\times \left[(\gamma-1)M+(\gamma+1)\sqrt{M^2-\frac{2
\eta_{2}\gamma q^2}{ (1+\gamma)}}\right]^{-1-\gamma}\label{th2}\;,\\
S(M,q)&=&\pi\gamma^{\gamma-1}\left(M+\sqrt{M^2-\frac{2
\eta_{2}\gamma q^2}{ (1+\gamma)}}\right)^{1+\gamma}\times\nonumber\\
&&\times \left[(\gamma-1)M+(\gamma+1)\sqrt{M^2-\frac{2
\eta_{2}\gamma q^2}{ (1+\gamma)}}\right]^{1-\gamma}\label{s2}\;,\\
A_{0}(M,q)&=&\frac{\eta_{2}}{\gamma q}\left(M-\sqrt{M^2-\frac{2\eta_{2}\gamma q^2}{(1+\gamma)}}\right)\;.\label{pot2}
\end{eqnarray}

Later on we will use the entropy as thermodynamic potential.  If we want to consider the mass, which is equivalent to the energy, as thermodynamic potential we need to invert (\ref{s2}) in order to write the mass in terms of the entropy and the electric charge. This is by no means evident from the general case (\ref{2}). Thus if we want to proceed with the analysis in terms of at least two different thermodynamic potential, i.e., entropy and mass, it is convenient to specialize to the case $\lambda=0$, $\gamma=1$ (with $\eta_1=\eta_2=1$), which is known as the Reissner-Nordstrom, or ($-\eta_1=\eta_2=-1$) which would be the anti-Reissner-Nordstrom. For these cases, the mass, electric charge, entropy and electric potential are given by
\begin{eqnarray}
\label{arn} \left\{\begin{array}{lr}
M=\frac{1}{2}\left(r_{+}+r_{-}\right)\; ,\quad q=\sqrt{\eta_2 r_{+}r_{-}}\;,\\
T=\frac{r_{+}-r_{-}}{4\pi r_{+}^2}\;,\quad S=\pi r_{+}^2\;,\quad A_{0}=\frac{q}{r_{+}}\;,
\end{array}\right.
\end{eqnarray}
and satisfy the first law of black hole thermodynamics (\ref{plt}) and the Smarr formula (\ref{smarr}).
\par
Finally, from (\ref{arn}) we have
\begin{equation}
r_{+}=M+\sqrt{M^2-\eta_{2}q^2}\;,\;r_{-}=M-\sqrt{M^2-\eta_{2}q^2}\label{r12}\;,
\end{equation}
which gives us the temperature, entropy and electric potential which explicitly read
\begin{eqnarray}
T&=&\frac{\sqrt{M^2-\eta_{2}q^2}}{2\pi\left(M+\sqrt{M^2-\eta_{2}q^2}\right)^2}\;,\label{th3}\\
S&=&\pi \left(M+\sqrt{M^2-\eta_{2}q^2}\right)^2\;,\;A_{0}=-\left(M-\sqrt{M^2-\eta_{2}q^2}\right)/q\;.\label{s3}
\end{eqnarray}

We have introduced all the necessary prerequisites to start with the study of the thermodynamic properties by means of the framework of Geometrothermodynamics. In the next section we will introduce this method and further apply it to our particular cases of interest.

\section{ The Geometrothermodynamics method}\label{sec3}

\hspace{0,6cm}The Geometrothermodynamics (GTD) make use of differential geometry as a tool to represent the thermodynamics of physical systems. Let us consider the $(2n+1)$-dimensional space $\mathbb{T}$, which coordinates are represented by the thermodynamic potential $\Phi$, the extensive variable $E^{a}$ and the intensive variables $I^{a}$, where $a=1,...,n$. If the space $\mathbb{T}$ has a non degenerate metric $G_{AB}(Z^{C})$, where $Z^{C}=\{ \Phi , E^{a} , I^{a}\}$, and the so called Gibbs 1-form  $\Theta=d\Phi -\delta_{ab}I^{a}dE^{b}$, with $\delta_{ab}$ the delta Kronecker; then the structure $\left(\mathbb{T} , \Theta, G\right)$ is said to be a contact riemannian manifold if $\Theta\wedge \left(d\Theta\right)^{n}\neq 0$ is satisfied \cite{hermann}. The space $\mathbb{T}$ is known as the thermodynamic phase space.
\par
We can define a $n$-dimensional subspace $\mathbb{E}\subset \mathbb{T}$, with extensive coordinates $E^{a}$, by the map $\varphi :\mathbb{E}\rightarrow\mathbb{T}$, with $\Phi\equiv \Phi (E^{a})$, such that
\begin{eqnarray}\label{cond1}
\varphi^{*}(\Theta)\equiv 0 \Rightarrow \left\{\begin{array}{ll}
d \Phi=\delta_{ab}I^{a}dE^{b}\;,\\
\frac{\partial\Phi}{\partial E^{a}}=\delta_{ab}I^{b}\;. \end{array}\right.
\end{eqnarray}
\par
We call the space $\mathbb{E}$ the thermodynamic space of the equilibrium states; the first equation (\ref{cond1}) is called ``First law of thermodynamics", and the second relation would be referred  as the ``condition of thermodynamic equilibrium". We impose as a necessary condition the ``Second law of thermodynamics":
\begin{equation}
\pm\frac{\partial^{2}\Phi}{\partial E^{a}\partial E^{b}}\geq 0\; ,\label{segunda}
\end{equation}
where the signal ($\pm$) depends on the chosen thermodynamic potential, for example, in the case of the mass we have ($+$), and for the case of entropy we have ($-$); this is known as convexity of the thermodynamic potential condition. The thermodynamic potential is defined such that it satisfy the homogeneity condition $\Phi (\alpha E^{a})=\alpha^{\beta}\Phi (E^{a})$. By differentiation with respect of $\alpha$, using the second equation of (\ref{cond1}) and taking $\alpha=1$, we have that
\begin{equation}
	\beta\Phi (E^{a})=\delta_{ab}I^{a}E^{b}\;.\label{euler}
\end{equation}

By differentiation of this last equation, and using the first equation of (\ref{cond1}), we have
\begin{equation}
	(1-\beta)\delta_{ab}I^{a} dE^{b}+\delta_{ab}E^{a} dI^{b}=0\; .\label{gibbsfor}
\end{equation}
For $\beta=1$ we get the Euler identity (\ref{euler}) and the Gibbs-Duhem formula (\ref{gibbsfor}). The pullback $\varphi^{*}:T^{*}(\mathbb{T})\otimes T^{*}(\mathbb{T})\rightarrow T^{*}(\mathbb{E})\otimes T^{*}(\mathbb{E})$\footnote{Where $T^{*}(\mathbb{T})$ and $T^{*}(\mathbb{E})$ represent the tangent spaces of $\mathbb{T}$ and $\mathbb{E}$, respectively.}, induce a metric on $\mathbb{E}$, such that $\varphi^{*}(G)=g$.

Hernando Quevedo has developed and improved a possible metric $G$ for the GTD, which for the case of black holes it can be written as \cite{quevedo}:
\begin{equation}
dL^{2}=G_{AB}dZ^{A}dZ^{B}=\Theta^{2}+\left(\delta_{ab}E^{a}I^{b}\right)\left(\eta_{ab}dE^{a}dI^{b}\right)\; \label{mt}\;,
\end{equation}
where $\eta_{ab} =\{\pm 1,1,...,1\}$. The case $\eta_{ab} =\{ -1,1,...,1\}$ holds for second order phase transition. The metric on $\mathbb{E}$ induced by the pullback $\varphi^{*}$ is:
\begin{align}
dl^{2} & =g_{ab}dE^{a}dE^{b}=\frac{\partial Z^{A}}{\partial E^{a}}\frac{\partial Z^{B}}{\partial E^{b}}G_{AB}dE^{a}dE^{b}\nonumber\\
 & =\left( E^{c}\frac{\partial\Phi}{\partial E^{c}}\right) \left( \eta_{ad}\delta^{di}\frac{\partial^{2}\Phi}{\partial E^{i}E^{b}}\right)dE^{a}dE^{b}\; . \label{me}
\end{align}

Since the thermodynamics system has to be independent of the particular choice of thermodynamic potentials, and also invariant under Legendre transformation, we have that the metrics (\ref{mt}) and (\ref{me}) should be invariant under Legendre transformations of the form
\begin{align}
\Phi=\widehat{\Phi}-\delta_{ab}\widehat{E}^{a}\widehat{I}^{b}\; , \; E^{a}=-\widehat{I}^{a}\; , \; I^{b}=\widehat{E}^{b}\; .\label{tl}
\end{align}
The n-dimensional space $\mathbb{E}$, with metric $g_{ab}$, contains the information about the thermodynamic interaction, phase transitions and fluctuations or stability of the thermodynamic system. With the help of metric (\ref{me}), we can derive the scalar curvature $R$, which gives us information about two aspects of the theory: when there are thermodynamic interaction and when there are phase transitions, also it tell us at which points of the thermodynamic equilibrium space these transitions take place.
\par
As we mentioned in the introduction, this method presents some important results. The first one is that the use of the Weinhold and Ruppeiner metrics provide results which are in contradiction to each other, even in contradiction to themselves when, for example, different thermodynamic potentials are used to describe the same system. We can cite the case of the Reissner-Nordstrom (RN) black hole for which the use of the Ruppeiner metric in the entropy representation provides a flat space $\mathbb{E}$, hence without phase transition \cite{aman}. However, when the representation is given by the internal energy of the black hole, the same method points out a non-zero scalar curvature with a singularity, i.e., with a phase transition in the curved space $\mathbb{E}$ \cite{shen}. This contradictory result is resolved in the GTD by the metric (\ref{mt}), which is invariant under Legendre transforms (\ref{tl}), reconciling the results of RN black hole thermodynamics regardless the choice of thermodynamic potential \cite{quevedo3}. The second result is the agreement with the usual analysis of the thermodynamic system by means of the specific heat of the black hole \cite{quevedo5}. A third result is that the description of the thermodynamic system depends, in many cases, of the ensemble's  choice \cite{myung, wu}, because this choice leads to different specific heats. This problem is solved by the description of GTD, so it results in a consistent description of the thermodynamic system described by different ensembles \cite{quevedo4}.    
\par
We saw that the metric of the thermodynamic equilibrium space  $\mathbb{E}$, can be obtained by the pullback of the metric defined on the contact riemannian space. By the definition of the line element (\ref{me}) in $\mathbb{E}$ space, we can define the distribution of probabilities to get a physical state with extensive variable $E^{a}$ within the interval $E^{a}+dE^{a}$:
\begin{align}\label{prob}
P(E^{a})=\frac{\sqrt{\det\left[g_{ab}\right]}}{(2\pi)^{\frac{n}{2}}}\exp\left[\frac{1}{2}g_{ab}dE^{a}dE^{b}\right]\; ,
\end{align}
where $P(E^{a})$ satisfy
\begin{align}
\int\prod\limits^{n}_{a=1}dE^{a}P(E^{a})=1\; .
\end{align}

It can be shown that by taking the derivative of (\ref{prob}) with respect of $V^{-1}$, where $V$ is the volume of the system, we obtain the expression  for the second fluctuations (in the thermodynamic limit $V\rightarrow\infty$) \cite{ruppeiner2}
\begin{align}
\langle \Delta E^{a}\Delta E^{b}\rangle =-g^{ab}\; ,\label{flut}
\end{align}
where $\Delta E^{a}=E^{a}-E^{a}_{(0)}$, and $g ^{ab}$ is the inverse of $g_{ab}$. A more realistic analysis requires the constants to be adjusted. If the fluctuations are small and real valued, then we say the system is stable.
\par
Another criteria used to determine the stability is trough the following geometric objects
\begin{eqnarray}
p_{1}^{(1)}&=&g_{11}>0\; ,\;p_{1}^{(2)}=g_{22}>0\; ,\, ...\, ,\; p_{1}^{(n)}=g_{nn}>0\; ,
\end{eqnarray}
\begin{eqnarray}
p_{2}^{(1)}&=&\left| \begin{array}{cc}  g_{11} & g_{12} \\
 g_{12} & g_{22}\end{array}\right|>0 \; ,\; p_{2}^{(2)}=\left| \begin{array}{cc}  g_{22} & g_{23} \\
 g_{23} & g_{33}\end{array}\right|>0\; ,\\
p_{3}^{(1)}&=&\left| \begin{array}{ccc}  g_{11} & g_{12}&g_{13} \\
 g_{12} & g_{22}&g_{23}\\
 g_{13} & g_{23}&g_{33}\end{array}\right| >0\; , \; p_{n}=\det\left[ g_{ab}\right]>0\; .
\end{eqnarray}
The positive (negative) signal  of $p_{n}$, which would depend on the choice of the thermodynamic potential, determines the local (un)stability of the thermodynamic system; more specifically if we have
\begin{align}
p_{i}>0\; ,\; i=1,\, ...\, ,n\; ,\label{eg}
\end{align}
then we can affirm the system is globally stable.
\par
Since we are not considering rotating black holes, another way to determine the global (un)stability is through the Helmholtz free energy. In terms of black hole thermodynamics variables, the Helmholtz free energy is no other thing than the Legendre transformation of mass (energy) $M(S,q)$:
\begin{align}
F(T,q)=M(S,q)-TS\;.\label{fh}
\end{align}
When we have
\begin{align}
F(E^{a})<0\; ,\; \forall E^{a}\; (E^{a}\in I(E^{a}))\; , \label{egh}
\end{align}
where $I(E^{a})$ is an interval and $E^{a}$ are the extensive variables, then the thermodynamic system is said to be globally stable. In the case of  usual thermodynamics the Helmholtz free energy is given by $F(T,V)=U-TS$.
\par
We can also define the Gibbs potential as
\begin{equation}
G(T,A_{0})=M(S,q)-TS-\eta_{2}A_{0}q\label{gibbs}\;,
\end{equation}
in this case the global stability is determined by
\begin{align}
G(E^{a})<0\; ,\; \forall E^{a}\; (E^{a}\in I(E^{a}))\; . \label{eggibbs}
\end{align}
We introduced the signal $\eta_{2}$ in (\ref{gibbs}) to compensate the contribution of the work term. Later we will make use of the Gibbs potential to determine the global stability of the thermodynamic system. For a system shown
stable in this ensemble, we must have
\begin{eqnarray}
\frac{\partial^2 G}{\partial T^2}\,,\frac{\partial^2 G}{\partial A_{0}^2}\,,\frac{\partial^2 G}{\partial T\partial A_{0}}\leq 0 \label{condg}\;.
\end{eqnarray}

In the next section we will apply these methods to determine the thermodynamic properties of the black hole solutions coming from the EMD theory.

\section{ Thermodynamics of phantom black holes}\label{sec4}

\subsection{ Application of the Geometrothermodynamics method}\label{sbsec4.1}

\subsubsection{ Einstein-(anti)Maxwell-Dilaton solutions}

\hspace{0,6cm}To begin with, let us define the thermodynamic variables of the system. For the Einstein-Maxwell-Dilaton black holes, we will have always a solution with two physical parameters, the mass $M$ and the electric charge $q$. Other variables (parameters) such as entropy $S$, temperature $T$ and electric potential $A_{0}$, can be defined as implicit functions of the parameters mentioned before. The contact riemannian manifold $\mathbb{T}$ is, in this case, $5$-dimensional and the space $\mathbb{E}$ of thermodynamic equilibrium states  is a 2-dimensional manifold.
\par
The thermodynamic description is the entropy representation, $S(M,q)$, which is identified as the thermodynamic potential $\Phi$, according to was defined in the previous section. The extensive variables are the mass $M$ and the electric charge $q$, which are represented by the coordinates $E^{a}$. The intensive variables are the temperature $T$ and the electric potential $A_{0}$, represented by the coordinates $I^{a}$.
\par
We have then a coordinate system for the thermodynamic phase space $\mathbb{T}$ as being $Z^{A}=\{ S(M,q), M, q, T, A_{0}\}$, together with the Gibbs 1-form given by \footnote{This expression comes from the first law of thermodynamics (\ref{plt}), which was inverted in order to isolate the entropy.}
\begin{equation}
\Theta_{S}=dS-\frac{1}{T}dM+\frac{\eta_{2}A_{0}}{T}dq\;,\label{fgs}
\end{equation}
such that $\varphi^{*}(\Theta_{S})=0$ is satisfied, which is no other thing than first law of black hole thermodynamics: $dM=TdS+\eta_{2} A_{0}dq$ ($\eta_{2} =\pm 1$).
\par
For a second-order phase transition, the line element (\ref{mt}) of the space $\mathbb{T}$ is
\begin{eqnarray}
dL^{2}&=&\left(dS-\frac{1}{T}dM+\frac{\eta_{2}A_{0}}{T}dq\right)^{2}+\left(\frac{M}{T}- \frac{\eta_{2}A_{0}}{T}q\right)\times\nonumber\\
&&\times\left[-d\left(\frac{1}{T}\right)dM+d\left(-\frac{\eta_{2}A_{0}}{T}\right)dq\right]\label{mt1}\;.
\end{eqnarray}
The first and second laws, and the equations of the equilibrium state as are given by
\begin{align}
d\Phi & =\delta_{ab}I^{a}dE^{b}\rightarrow dS=\frac{1}{T}dM-\frac{\eta_{2} A_{0}}{T}dq\;,\\
& \frac{\partial^{2} S}{\partial M^{2}}\,, \frac{\partial^{2} S}{\partial M \partial q}\, , \frac{\partial^{2} S}{\partial q^{2}}\leqslant 0\;,\\
\frac{\partial \Phi}{\partial E^{a}} & =\delta_{ab}I^{b}\rightarrow \frac{\partial S}{\partial M}=\frac{1}{T}\, ,\, \frac{\partial S}{\partial q}=-\frac{\eta_{2} A_{0}}{T}\; .
\end{align}

Now we have to specify a solution, where we consider first the general case (\ref{2}). The line element (\ref{me}) of the equilibrium space, taking into account (\ref{s2}), would be
\begin{align}
dl^{2}&=\left(M\frac{\partial S}{\partial M}+q\frac{\partial S}{\partial q}\right)\left(-\frac{\partial^2 S}{\partial M^2}dM^2+\frac{\partial^2 S}{\partial q^2}dq^2\right)\;,\label{me1}\\
&=g_{MM}dM^2+g_{qq}dq^2\label{marng},\\
g_{MM}&=16\pi^2\gamma^{2\gamma}\frac{\left(M+\sqrt{M^2-\frac{2\eta_{2}\gamma q^2}{(1+\gamma)}}\right)^{1+2\gamma}}{ \sqrt{M^2-\frac{2\eta_{2}\gamma q^2}{(1+\gamma)}}}\times\nonumber\\
&\quad\times\left(M^2-\eta_{2}q^2+M\sqrt{M^2-\frac{2\eta_{2}\gamma q^2}{(1+\gamma)}}\right)\times\nonumber\\
&\quad\times\left[(\gamma-1)M+(\gamma+1)\sqrt{M^2-\frac{2\eta_{2}\gamma q^2}{(1+\gamma)}}\right]^{-1-2\gamma}\times\nonumber\\
&\quad\times\left[(1+\gamma)M-(1+3\gamma)\sqrt{M^2-\frac{2\eta_{2}\gamma q^2}{(1+\gamma)}})\right]\;,\\
g_{qq}&=-32\eta_{2}\pi^2\gamma^{1+2\gamma}\frac{\left(M+\sqrt{M^2-\frac{2\eta_{2}\gamma q^2}{(1+\gamma)}}\right)^{-1+2\gamma}}{\sqrt{M^2-\frac{2\eta_{2}\gamma q^2}{(1+\gamma)}}}\times\nonumber\\
&\quad\times\left(M^2-\eta_{2}q^2+M\sqrt{M^2-\frac{2\eta_{2}\gamma q^2}{(1+\gamma)}}\right)\times\nonumber\\
\end{align}
\begin{align}
&\quad\times\left[(\gamma-1)M+(\gamma+1)\sqrt{M^2-\frac{2\eta_{2}\gamma q^2}{(1+\gamma)}}\right]^{-1-2\gamma}\times\nonumber\\
&\quad\times\left[M^3+(M^2-\eta_{2}q^2)\sqrt{M^2-\frac{2\eta_{2}\gamma q^2}{(1+\gamma)}}\right]\;.
\end{align}
We notice that  the phantom contribution can switch the signature of the metric of the space $\mathbb{E}$. To calculate the scalar curvature  associated with the metric (\ref{marng}), we can make use of a particular formula valid for 2-dimensional spaces
\begin{align}
R(M,q)&=-\frac{1}{\sqrt{|det[g]|}}\left[\partial_{q}\left(\frac{\partial_{q}g_{MM}-\partial_{M}g_{Mq}}{\sqrt{|det[g]|}}\right)+\partial_{M}\left(\frac{\partial_{M}g_{qq}-\partial_{q}g_{Mq}}{\sqrt{|det[g]|}}\right)\right]\nonumber\\
&\quad-\frac{det[H_{S}]}{2\left(det[g]\right)^2}\;,\label{ricci}\\
H_{S}&=\left(\begin{array}{rrr}
g_{MM}&g_{Mq}&g_{qq}\\
\partial_{M}g_{MM}&\partial_{M}g_{Mq}&\partial_{M}g_{qq}\\
\partial_{q}g_{MM}&\partial_{q}g_{Mq}&\partial_{q}g_{qq}
\end{array}\right)\;.\label{harng}
\end{align}
Replacing (\ref{marng}) into (\ref{ricci}), we have
\begin{equation}
R(M,q)=\frac{N(M,q)}{D(M,q)}\label{riccig}\,,
\end{equation}
where
\begin{align}
N(M,q)&=S_{MM}^2\left(FF_{q}S_{qqq}+2S_{qq}\left(F_{q}^2-FF_{qq}\right)\right)\nonumber\\
&\quad+FS_{qq}\left(-S_{qq}F_{M}S_{MMM}+F\left(S_{MMq}^2-S_{MMM}S_{qqq}\right)\right)\nonumber\\
&\quad+S_{MM}\Big(-2S_{qq}^2F_{M}^2+FS_{qq}\left(-F_{q}S_{MMq}+F_{M}S_{qqM}+2S_{qq}F_{MM}\right)\nonumber\\
&\quad+F^2\left(S_{MMq}S_{qqq}-S_{qqM}^2-2S_{qq}\left(S_{MMMM}-S_{qqMM}\right)\right)\Big)\;,\label{N}
\end{align}
with $\partial_{i}...\partial_{j}S=S_{i...j}$, $\partial_{i}...\partial_{j}F=F_{i...j}$, and
\begin{eqnarray}
D(M,q)=2F^3S_{MM}^2S_{qq}^2\;,\label{D}
\end{eqnarray}
where
\begin{equation}
F=MS_{M}+qS_{q}\label{F}\;,
\end{equation}
and S as given before by (\ref{s2}).
\par
With the help of a mathematical software, we find the zeros of the scalar curvature $R(M,q)$ (see (\ref{riccig})) are the mass values $M_{1}=\pm q\sqrt{2\eta_2\gamma/(1+\gamma)}$, $M_{2}=\pm q\sqrt{\eta_{2}(1+\gamma)/2}$. We also have another zero for the scalar curvature for the particular case $\eta_2=-1$  at $M_{3}=q\sqrt{(1+\gamma)/2(1+2\gamma)}$. This tell us that in general the scalar curvature is non zero - which implies the existence of thermodynamic interaction for this class of black holes \cite{quevedo2} - and is zero only when the mass takes the values $M_1$, $M_2$ or $M_3$. These values characterize when the the black hole becomes extremal. Let us note that only when we set $r_{+}=r_{-}$ in (\ref{r1})-(\ref{r2}), we obtain the value $M_2$ for the mass. Probably this is related to the fact we pointed out before, that in the normal and phantom cases, $r_{-}$ is a  true singularity of the curvature, therefore, the extremal limit in this regime is not viable. The two values $M_{1}$ and $M_2$, for $\eta_{2}=\gamma=1$ (RN), are $M_{1,2}=\pm q$, in total agreement with \cite{davies}. We also want to note that there exist an extremal limit at $M_3$, when $\eta_2=-1$. The point at which the scalar $R(M,q)$ diverge is given by the mass $M_{4}=\pm q(1+3\gamma)/\sqrt{2\eta_2 (1+\gamma)(1+2\gamma)}$. Then, when $\eta_{2}=\gamma=1$, $M_{4}=\pm 2q/\sqrt{3}$, also in agreement with the previous work of Davies \cite{davies}. However, for the anti-RN solution $M_4$ is not real, which means that there is no phase transition for the phantom when work with the entropy representation.
\par
To corroborate our analysis, let us calculate the specific using the well know formula
\begin{equation}
C_{q}=\left(\frac{\partial M}{\partial T}\right)_{q}=\left(\frac{\partial M}{\partial S}\right)_{q}\Big/\left(\frac{\partial^2 M}{\partial S^2}\right)_{q}=-\left(\frac{\partial S}{\partial M}\right)_{q}^2\Big/\left(\frac{\partial^2 S}{\partial M^2}\right)_{q}\label{cqs}\;.
\end{equation}
Doing this we obtain:
\begin{eqnarray}
C_q&=&4\pi\gamma^{\gamma}\sqrt{M^2-\frac{2\eta_{2}\gamma q^2}{(1+\gamma)}}\left(M+\sqrt{M^2-\frac{2\eta_{2}\gamma q^2}{(1+\gamma)}}\right)^{1+\gamma}\times\nonumber\\
&&\times\frac{\left[(\gamma -1)M+(1+\gamma)\sqrt{M^2-\frac{2\eta_{2}\gamma q^2}{(1+\gamma)}}\right]^{1-\gamma}}{(1+\gamma)M-(1+3\gamma)\sqrt{M^2-\frac{2\eta_{2}\gamma q^2}{(1+\gamma)}}}\;.\label{cqs1}
\end{eqnarray}

The zeros of (\ref{cqs1}) give us the information about the points  at which the black hole becomes extremal; these are precisely $M_1$ and $M_2$ founded  by the previous analysis, thus corroborating our analysis. In the same way, we get that the phase transition point at which $C_q$ diverges is $M_4$. Remarkably is the fact that $M_3$ is not  a zero of the specific heat, hence, it should be some spurious zero that results from a failure of applicability of the analysis for this class of pathological solutions. We will see below that not only a new (non physical) extremal case is revealed by the GTD method, but also a new critical point is revealed when we choose the  mass of the black hole as the thermodynamic potential.
\subsubsection{ (anti)Reissner-Nordstrom solution}

\hspace{0,6cm}To study in detail the thermodynamics of the EMD system, let us consider the particular case $\gamma=1$, this corresponds to the (anti)Reissner-Nordstrom ((anti)RN) solution. For this case, we have that the entropy is
\begin{equation}
S=\pi \left(M+\sqrt{M^2-\eta_{2}q^2}\right)^2\label{s3}\;,
\end{equation}
that,  when replaced the  into (\ref{marng}), furnish the following:
\begin{align}
dl^2&=-4\pi^2\frac{\left(M^2-\eta_{2}q^2+M\sqrt{M^2-\eta_{2}q^2}\right)}{\left(M^2-\eta_{2}q^2\right)^2}\times\nonumber\\
&\quad\times\Big\{\left(M+\sqrt{M^2-\eta_{2}q^2}\right)^3\left(-M+2\sqrt{M^2-\eta_{2}q^2}\right)dM^2\nonumber\\
&\quad+\eta_{2}\left(M+\sqrt{M^2-\eta_{2}q^2}\right)\left[\left(M^2-
\eta_{2}q^2\right)^{3/2}+M^3\right]dq^2\Big\}\label{mrnarn}\;.
\end{align}

The scalar curvature derived from  this  metric is given by (\ref{riccig}), where
\begin{align}
N(M,q)&=-\sqrt{M^2-\eta_{2}q^2}\left(M+\sqrt{M^2-\eta_{2}q^2}\right)
\Big[1024M^{14}-3840\eta_{2}q^{2}M^{12}\nonumber\\
&\quad+5504q^{4}M^{10}-3408\eta_{2}q^{6}M^{8}+
460q^{8}M^{6}+377\eta_{2}q^{10}M^{4}\nonumber\\
&\quad-122q^{12}M^{2}+5\eta_{2}q^{14}+\sqrt{M^2-\eta_{2}q^2}
\Big(  1024m^{13}\nonumber\\
&\quad-3328\eta_{2}q^{2}M^{11}+3968q^{4}M^{9}
-1776\eta_{2}q^{6}M^{7}-100q^{8}M^{5}\nonumber\\
&\quad+251\eta_{2} q^{10}M^{3}-35q^{12}M\Big)\Big]\label{N1}\;,\\
D(M,q)&=4\pi^2\Big[ 64M^{10}-208\eta_{2}q^{2}M^{8}+248q^{4}M^{6}-137\eta_{2}q^{6}M^{4}\nonumber\\
&\quad+34q^{8}M^{2}-2\eta_{2}q^{10}+\sqrt{M^{2}-\eta_{2}q^2}\times\Big( 64M^{9}-176\eta_{2}q^{2}M^{7}\nonumber\\
&\quad+168q^{4}M^{5}-71\eta_{2}q^{6}M^{3}+11q^{8}M\Big)\Big]^2\label{D1}\;.
\end{align}

The zeros of the numerator, $N(M,q)$, in (\ref{N1}), are $M_5=\pm q\sqrt{\eta_2}$ and $M_6=\pm iq\sqrt{\eta_2/3}$, which depend on the choice of $\eta_2$. The RN case, when $\eta_{2}=1$, has the zero $M_5=\pm q$ (where we did $r_+=r_-$) that corresponds to the extremal RN black hole. The anti-RN case, $\eta_2=-1$, has the zero $M_6=\pm q/\sqrt{3}$, but this result reveals a weakness of this method, because it points out the presence  of an extremal anti-RN black hole, which we know does not exist. The analysis of the causal structure performed in \cite{gerard1} shows that the anti-RN black hole has a causal structure identical to the Schwarzschild black hole, hence, there is no exist extremal limit for this case. We don't have any explanation for such irregularity of the thermodynamic system when described by the GTD method. We believe that the extremal limit for non trivial black holes, just like the anti-RN case, reveals pathologies which are not well described by the GTD.
\par
The zero of the denominator $D(M,q)$, in (\ref{D1}), is given by $M_7=\pm 2q\sqrt{\eta_2/3}$, that is real only for $\eta_2=1$. It occurs then that the RN black hole has a second order phase transition point at $M_7$, in good correspondence with Davies \cite{davies}. On the other hand, the anti-RN black hole, $\eta_{2}=-1$, has not any phase transition point; consequently, there is not extremal analogue, nor phase transition for the anti-RN case.
\par
As before, order to check our results, let as can calculate the specific heat of these model. The specific heat (\ref{cqs}), calculated in combination with (\ref{s3}), is \begin{align}
C_q&=2\pi\sqrt{M^2-\eta_{2}q^2}\frac{\left(M+\sqrt{M^2-\eta_{2}q^2}\right)^2}{\left(M-2\sqrt{M^2-\eta_{2}q^2}\right)}\label{ce1}\;.
\end{align}
From (\ref{ce1}) we have that the black hole is extremal ($C_q=0$) when $M=M_5=\pm q\sqrt{\eta_2}$, and it has a point of phase transition ($C_q\rightarrow\infty$) at $M=M_7=\pm 2q\sqrt{\eta_2/3}$ . These results are in total agreement with the previous results we found by the GTD method, but we want to stress that before we obtained an extra  extremal anti-RN black hole solution. This reinforce our previous statement about the weakness of the GTD method for some pathological situations.
\par
As was mentioned before, we will consider now the mass representation analysis. We start by inverting (\ref{s3}) to write the mass in terms of the entropy and electric charge, this is
\begin{equation}
M(S,q)=\frac{S+\eta_{2}\pi q^2}{2\sqrt{\pi S}}\label{m1}\;.
\end{equation}
The Gibbs 1-form reads in this case
\begin{equation}
\Theta_{M}=dM-TdS-\eta_{2}A_{0}dq\label{gibbsm}\;.
\end{equation}
Then, using  (\ref{mt}) we obtain
\begin{eqnarray}
dL^2=\left(dM-TdS-\eta_{2}A_{0}dq\right)^2+\left(TS-\eta_{2}A_{0}q\right)\left[-dTdS+d\left(-\eta_{2}A_{0}\right)dq\right]\label{mt2}.
\end{eqnarray}

The pullback $\varphi^{*}$ induce a metric on the space $\mathbb{E}$,
\begin{align}
dl^2&=\left(S\frac{\partial M}{\partial S}+q\frac{\partial M}{\partial q}\right)\left(-\frac{\partial^2 M}{\partial S^2}dS^2+\frac{\partial^2 M}{\partial q^2}dq^2\right)\label{me2}\\
&=\frac{(S+3\eta_{2}\pi q^2)}{4S}\left[\frac{(S-3\eta_{2}\pi q^2)}{8\pi S^2}dS^2+\eta_{2}dq^2\right]\label{me3}\;.
\end{align}
Finally, with this metric we can compute the scalar of curvature,
\begin{eqnarray}
R(S,q)=\eta_{2}\frac{288\pi^2 q^2 S^2(S-\eta_{2}\pi q^2)}{(S-3\eta_{2}\pi q^2)^2(S+3\eta_{2}\pi q^2)^3}\label{ricci3}\;.
\end{eqnarray}

We can see that when  the thermodynamic potential is the mass of the RN black hole,  the interpretation by this method of the extremal limit  ($r_+=r_-$)  is clear: the zeros of the numerator of $R(S,q)$ in (\ref{ricci3})  exist only for the values  $S_1=\pi q^2$ , which clearly represent the extremal RN black hole. When the anti-RN (phantom) case, $\eta_{2}=-1$ , is considered, we evidence that there is no extremal limit, as it should be. This last conclusion solve in some sense the unphysical prediction we get in the entropy representation, by changing to the mass  representation. But in contrast a new problem arise, we get a new point of phase transition for the anti-RN case at $S_2=3\eta_{2}\pi q^2$, with $r_+=3\eta_{2}r_-$. This indicates a breach of the invariance of the theory for describe the system regardless of choice of thermodynamic potential. Once again this result confirms our claim about the failure of the GTD method when applied to pathological solutions, which despite having a simple causal structure, apparently they present non trivial solutions. For the RN solution we have a extremal limit at $S_1$ and a second order phase transition point at $S_2$, in agreement with \cite{davies}.
\par
Finally, let us compute the specific heat. Using (\ref{cqs}) together with (\ref{m1}), we get
\begin{equation}
C_q=-2S\left(\frac{S-\eta_{2}\pi q^2}{S-3\eta_{2}\pi q^2}\right)\label{cqm}\;.
\end{equation}
From this formula  the correct interpretation can be read: for the RN black hole we have an extremal limit when $S=S_1=\pi q^2$, and a phase transition point at $S=S_2=3\pi q^2$. On the other hand, for the anti-RN case, we conclude there is no extremal limit, nor point of phase transition.

\subsection{Local and global stability}\label{sbsec4.2}


\hspace{0,6cm}It is usual to study local stability of a thermodynamic system by means of the specific heat. Alternatively, within the GTD, we can study the metric components of the thermodynamic equilibrium space $\mathbb{E}$ or even more, the Hessians of the entropy and mass. To determine the global stability it can be achieved by the analysis of all the components of the metric as well as their corresponding determinants; but also by means of the Helmholtz free energy or by the Gibbs potential. Here we will continue with the study of local and global stability of EMD black holes solutions.
\par
Let us start calculating the Hessian  of the entropy for the general case (\ref{s2}), which is defined as
\begin{eqnarray}\label{HS}
H_{S}=\left(\begin{array}{cc}
\frac{\partial^2 S}{\partial M^2}&\frac{\partial^2 S}{\partial M\partial q}\\
\frac{\partial^2 S}{\partial M\partial q}&\frac{\partial^2 S}{\partial q^2}
\end{array}\right)\;.
\end{eqnarray}
Using (\ref{s2}) and considering $\eta_2=1$ (including $\gamma=1$) we get local instability in view of $S_{MM}(M,q)$ is always positive, and $S_{qq}(M,q)$ and $S_{Mq}(M,q)$ are positive as long as $q>0$. In the same manner, when $\eta_2=-1$, $S_{MM}(M,q)$ and $S_{qq}(M,q)$ are always positive and $S_{Mq}(M,q)$ is  positive provided that $q>0$; therefore we have that the solutions are locally unstable.
\par
Analogously, we can calculate the Hessian matrix of the mass for the particular choice $\gamma=1$. Using  (\ref{m1}) we found that
\begin{align}\label{HS}
H_{M}&=\left(\begin{array}{cc}
\frac{\partial^2 M}{\partial S^2}&\frac{\partial^2 M}{\partial S\partial q}\\
\frac{\partial^2 M}{\partial S\partial q}&\frac{\partial^2 M}{\partial q^2}
\end{array}\right)\\
&=\left(\begin{array}{cc}
\frac{3\eta_2\pi q^2-S}{8\sqrt{\pi}S^{5/2}}&-\frac{\sqrt{\pi}\eta_2 q}{2S^{3/2}}\\
-\frac{\sqrt{\pi}\eta_2 q}{2S^{3/2}}&\frac{\eta_2 \sqrt{\pi}}{S^{1/2}}
\end{array}\right)\;,
\end{align}
The Hessian of the mass also leads us to local instabilities. More precisely, when $\eta_2=1$, then $M_{SS}$ and $M_{Sq}$ can take positive or negative values. Similarly, when $\eta_2=-1$, $M_{Sq}$ can takes  positive or negative value, whereas $M_{SS}$ and $M_{qq}$ are always negative.
\par
Once again, let us consider the analysis by the study of the specific heat (\ref{cqs1}), which in terms of $r_+$ and $r_-$ can be written as
\begin{eqnarray}
C_q=-2\pi r_+^{1+\gamma}\frac{(r_+-\gamma r_-)(r_+-r_-)^{1-\gamma}}{\left[r_+-(1+2\gamma)r_-\right]}\label{cqr}\;.
\end{eqnarray}

From this expression, we can establish, for example,  local stability ($C_q>0$) for the event horizon interval $\gamma r_-<r_+<(1+2\gamma)r_-$, $0<\gamma<1$ (EMD with $\eta_1=1$). The specific heat, with $\gamma=\eta_2=1$ or $\gamma=-\eta_2=1$, corresponds to the RN or anti-RN cases, respectively. The anti-RN case with $\gamma=1$ is locally  unstable because $r_-<0$ implies $C_q<0$. The only case where the phantom solutions are locally stable will be when $\gamma r_-<r_+<(1+2\gamma)r_-$ for $\gamma<-1$.
\par
As we mentioned before, the components of the metric of the thermodynamic equilibrium spac, $\mathbb{E}$, give us also information about local stability. According to (\ref{me1}), the components of $g_{MM}$ and $g_{qq}$, written in terms of $r_{+}$ and $r_-$, read
\begin{align}
g_{MM}&=-16\pi^2r_{+}^{2(1+\gamma)}\frac{\left[r_{+}-(1+2\gamma)r_{-}\right]}{(r_{+}-\gamma r_{-})(r_{+}-r_{-})^{2\gamma}}\;,\\
g_{qq}&=-8\eta_{2}\pi^2r_{+}^{1+2\gamma}\frac{\left[r_{+}^2+
(\gamma-1)r_{-}+2\gamma (\gamma+\frac{1}{2})r_{-}^2\right]}{(r_{+}-\gamma r_{-})(r_{+}-r_{-})^{2\gamma}}\;.
\end{align}
These components always have opposite signals for each of the two cases, normal and phantom, including the case $\gamma=1$. Therefore, by simple inspection of the components of the metric on $\mathbb{E}$, we assert  that the system is locally unstable; additionally, it  tells us that there is global instability, just as was mentioned in Section $3$.
\par
For the next step, let us study the global stability of the class of solutions of interested. For the general case we can write
\begin{eqnarray}
M&=&\frac{q}{2A_{0}}\left[1+\frac{2\eta_{2}\gamma A_{0}^2}{(1+\gamma)}\right]\;,\;TS=\frac{q}{4A_{0}}\left[1-\frac{2\eta_{2} A_{0}^2}{(1+\gamma)}\right]\;,\label{eq1}\\
r_{+}&=&\frac{1}{4\pi T}\left[1-\frac{2\eta_{2}A_{0}^2}{(1+\gamma)}\right]\;.\label{eq2}
\end{eqnarray}
The analysis start with the grand canonical ensemble. Using (\ref{eq1}) and (\ref{eq2}), we calculate the Gibbs potential (\ref{gibbs}):
\begin{equation}
G(T,A_{0})=\frac{1}{16\pi T}\left[1-\frac{2\eta_{2}A_{0}^2}{(1+\gamma)}\right]^{1+\gamma}\label{gibbs1}\;.
\end{equation}

We know that $T>0$, so we have to analyze the term between brackets. Consider first the normal case with $\eta_{2}=1$; when $(1+\gamma)$ is an odd integer, then the system is globally stable only for $\gamma>0$ and $A_{0}\in(-\infty,-\sqrt{(1+\gamma)/2})$ $\cup$ $ (\sqrt{(1+\gamma)/2},+\infty)$. The critical electric potential values are $A_{0\,(c)}=\pm\sqrt{(1+\gamma)}$ $(\sqrt{2}^{-1})$ (by setting $r_{+}=0$ in (\ref{eq2})). No let us consider the phantom solutions with $\eta_{2}=-1$, in this case the system is globally unstable, except for $\gamma <-1$ and $A_{0}\in(-\infty,-\sqrt{|1+\gamma|/2})$ $\cup$ $ (\sqrt{|1+\gamma|/2},+\infty)$. A straightforward calculation of the derivatives in (\ref{condg}) shows that
\begin{align}
\frac{\partial^2 G}{\partial T^2}&=\frac{1}{8\pi T^3}\left[1-\frac{2\eta_2 A_{0}^2}{(1+\gamma)}\right]^{1+\gamma}\;,\\
\frac{\partial^2 G}{\partial A_{0}^2}&=-\eta_2\frac{(1-2\eta_2 A_{0}^2)}{4\pi T}\left[1-\frac{2\eta_2 A_{0}^2}{(1+\gamma)}\right]^{\gamma-1}\;,\\
\frac{\partial^2 G}{\partial T\partial A_{0}}&=\frac{\eta_2 A_{0}}{4\pi T^2}\left[1-\frac{2\eta_2 A_{0}^2}{(1+\gamma)}\right]^{\gamma}\label{condg1}\;.
\end{align}

From here we can see that (\ref{condg1}) always break the stability criteria, because $A_{0}$ can be positive or negative (remember that $q$ is real). Then it follows that the system is always locally unstable.
\par
The corresponding analysis by the canonical ensemble formalism is slightly different. We start writing the Helmholtz free energy (\ref{fh}):
\begin{eqnarray}
F=\frac{1}{16\pi T}\left[1-\frac{2\eta_2 A_{0}^2}{(1+\gamma)}\right]^{\gamma}\left[1+2\eta_2 A_{0}^2\left(\frac{1+2\gamma}{1+\gamma}\right)\right]\label{fh1}\;.
\end{eqnarray}
The analysis of this formula is the following: the system is globally stable for two cases. The first ones when $A_{0}\in(-\infty,-\sqrt{(1+\gamma)/2})\cup(\sqrt{(1+\gamma)/2},+\infty)$ for $\gamma>0$ odd integer and $\eta_{2}=1$. The second case is when $A_{0}\in(-\infty,$ $-\sqrt{(1+\gamma)/2(1+2\gamma)})$ $\cup$ $(\sqrt{(1+\gamma)/2(1+2\gamma)},+\infty)$ for $\gamma$ even integer and $\eta_{2}=-1$ or $A_{0}\in(-\infty,-\sqrt{(1+\gamma)/2})$ $\cup$  $(\sqrt{(1+\gamma)/2},+\infty)$ for $\gamma\in\Re$. When  we use the specific heat formula (\ref{cqr}), we found second order phase transition points given by $A_{0\,(1)}=\pm\sqrt{\eta_2 (1+\gamma)/2(1+2\gamma)}$ and $A_{0\,(2)}=\pm\sqrt{\eta_2 (1+\gamma)/2}$ ( even for $\gamma>1$) . The point $A_{0\,(3)}=\pm\sqrt{\eta_2 (1+\gamma)/(2\gamma)}$, as well as  $A_{0\,(2)}$ with $\gamma<1$, represent the extremal case where $r_{+}\rightarrow 0$ (see \cite{gerard1} for details).
\par
Finally, we can calculate the minimum temperature for this system using the formula $\partial T^{-1}/\partial r_+=0$ \cite{rabin}, so we have:
\begin{eqnarray}
T_0=\frac{1}{4\pi r_+}\left(\frac{\gamma}{1+\gamma}\right)^{\gamma}\label{t0}\;.
\end{eqnarray}
From the specific heat, Eq. (\ref{cqr}), we identify a phase transition point at $r_+=3r_-$, which gives us the critical temperature
\begin{eqnarray}
T_c=\frac{1}{4\pi r_+}\left(\frac{2\gamma}{1+2\gamma}\right)^{\gamma}\label{tc}\;,
\end{eqnarray}
combining this with (\ref{t0}) we get the relation $T_c=T_0 [2(1+\gamma)/(1+2\gamma)]^{\gamma}$.

\section{Conclusions}\label{sec5}
The thermodynamic properties of the class of solutions known as phantom black holes has not been studied in detail yet. The physical stability of these solutions can be determined also by studying their thermodynamic properties. The main objective of the present work was to fill this gap by establishing a detailed analysis of this kind of solutions.
\par
The zeroth, second and third laws of thermodynamics remain unmodified by these solutions. However, the first law had to be generalized  to take into account  the contribution of the work done on (or by) the system with the ``wrong" sign (when compared to the usual case). The differential form of the generalized first law was written in (\ref{plt}) whereas the exact expression was written in  (\ref{smarr}).
\par
The use of Geometrothermodynamics as a tool of analysis has proven to be, once more,  equivalent to the most usual methods, but with an important exception that was the phantom case with $\eta_2=-1$. More specifically,  we saw for the phantom solutions that when we choose the entropy as the thermodynamic potential, the method introduce a new value of the mass parameter which could be interpreted as a (non physical) extremal black hole limit. If we choose the the mass as the thermodynamic potential, then we get a new critical point for the system. The new mass parameter and the new critical point have to be considered as spurious zeros of the numerator and demininador of the scalar curvature of the space $\mathbb{E}$. This can be understood  by the fact that EMD solutions switch to the phantom sector (i.e. with $\eta_2=-1$) by the symmetry transformation $q^2\rightarrow -q^2$, as was shown in \cite{gerard2}. In the case of the entropy representation, with the choice $\gamma=1$, that kind of symmetry leads to the appearance of a new real valued zero of the scalar curvature which  originally was  pure imaginary. In the case of the mass representation, the same symmetry transformation entails the appearance of a new critical point, i.e., a divergence of the scalar curvature.  When we carried out the study of the thermodynamic properties by means of specific heat,  we have not found new spurious critical points nor new divergence points of the scalar curvature, just as was expected.  This result revealed the fragility of the GTD method when applied to pathological solutions.
\par
In regard to the stability analysis, we found that the only possible local stability would correspond to the following  cases: $\gamma r_{-}<r_{+}<(1+2\gamma)r_-$, for $\eta_1=\eta_2=1$ and $0<\gamma<1$ or for $\eta_{1}=\eta_{2}=-1$ and $\gamma<-1$. All the other solutions, normal or phantom, has been shown to be locally unstable. The global stability can be established for some particular situations that restrict the values the electric potential $A_0$ and the parameter $\gamma$ can take.
\par    
The perspectives of the present work is to study in detail the subtleties that arise from the phantom solutions, with the hope to strength some weak aspects of the promising novel  method which proved to be the Geometrothermodynamics. But also we expect that will be evidenced some physical limits to the use of this method. This has already been shown to be true in the cases of Reissner-Nordstrom-AdS (RN-AdS) and anti-RN-AdS black holes \cite{deborah}. We propose elucidate with more details this issue in a forthcoming work.
\par
Finally, with respect to the global stability of one portion of the class of phantom solutions, we expect this to be an indication for the stability of the space-time of these class of solutions. That shall also be a topic to be studied in a subsequent work. 

\vspace{0,25cm}
{\bf Acknowledgement:}\,We are grateful to Gabriela Conde Saavedra for the help in the elaboration of the manuscript. M. E. Rodrigues  thanks  UFES for the hospitality during the development of this work and Z. Oporto thanks  CLAF/CNPq for financial support.


\end{document}